\documentclass[prd,aps,showpacs]{revtex4}
\usepackage{epsfig}

\def\vk{{\bf k}}
\def\mg{{\big <}}
\def\md{{\big >}}
\def\mP{{\cal P}}
\def\mB{{\cal B}}
\def\d{{\bf d}}
\def\deltal{{\delta_{\rm lin.}}}
\newcommand{\dd}{{\rm d}}
\begin{document}
\begin{flushright}
  SPhT-Saclay t04/110
\end{flushright}

\title{Constraints on higher-dimensional gravity from the cosmic shear three-point correlation function}
\author{Francis Bernardeau}
\affiliation{Service de Physique Th{\'e}orique,
             CEA/DSM/SPhT, Unit{\'e} de recherche associ{\'e}e au CNRS,
             CEA/Saclay F-91191 Gif-sur-Yvette cedex, France.}

\date{\today}

\begin{abstract}
With the developments of large galaxy surveys or cosmic shear
surveys it is now possible to map the dark matter distribution at
truly cosmological scales. Detailed examinations of the
statistical properties of the dark matter distribution reveal the
detail of the large-scale structure growth of the Universe. In
particular it is shown here that the behavior of the density field
bi-spectrum is sensitive to departure from normal gravity in a way
which depends only weakly on the background evolution. The cosmic
shear bispectrum appears to be particularly sensitive to changes
in the Poisson equation: we show that the current cosmic shear
data can already be used to infer constraints on the scale of a
possible higher-dimensional gravity, above 2 $h^{-1}$Mpc.
\end{abstract}
 \maketitle

\vskip2pc

%%%%%%%%%%%%%%%%%%%%%%%%%%%%%%%%%%%%%%%%%%%%%%%%%%%%%%%%%%%%%%%%%%%

Almost all cosmological observations point toward a coherent model
in which there is a non-vanishing dark energy~\cite{wmap}. The
existence of such a component is very puzzling from a high energy
physics point of view and signals physics beyond the standard
model in way we do not yet apprehend. The resolution of such an
issue may lie either in the introduction of a new form of cosmic
fluid - a dark energy -  with a negative pressure that would
correspond to two third of the actual energy density of the
universe or maybe in a modification of the Einstein gravity
equations at cosmological scales. The two scenarios have obviously
to be distinguished~\cite{diffDE}! The first route is usually
taken in models evoking quintessence component. However it cannot
excluded that a non-zero cosmological constant observation
actually signals a modification of the Einstein law at
cosmological scales. This idea is actually fuelled by recent
developments in string theory phenomenologies. Indeed the
introduction of branes in the context of superstring
theories\cite{RS,BDL}, led to concepts of higher dimensional
spacetimes. Usually it is assumed that the interaction gauge
fields are localized on a 3--brane (i.e. a 3+1 dimension
spacetime) whereas gravity propagates in all dimensions. In any of
such string inspired models, one expects both the existence of
Kaluza-Klein gravitons implying a non standard gravity on very
small scales and light bosons, which can manifest as a new
fundamental small scale force. It cannot be however excluded that
gravity might be modified at large cosmological scale as well
~\cite{GRS,kogan,largedim,RSgrav}. This possibility has been
explicitly linked to the evidences of a cosmological
constant~\cite{Deffayet1,Deffayet2}. It asks for precise test of
gravity law at cosmological scales. This is the aim of this letter
in which we pursue investigations on how cosmological
observations, namely cosmic shear observations, could be used to
test gravity on large distances.

Deviations form Newtonian or Einstein gravity have been looked for
in the past mainly at rather small scales (from a cosmological
point of view) whether it is in laboratory
experiments~\cite{testgrav1,submm} or in stellar systems. That put
severe constraints on the post-Newtonian
parameters~\cite{testgrav2,damour} but tells little on the
validity of the Einstein equations at cosmological scales (say at
scales above $1\,h^{-1}$Mpc). At best one can invoke studies of
galaxy clusters. Comparisons between X--ray emissivity and
gravitational lensing, which appears to be an indirect test of the
Newton law through the equation of hydrostatic equilibrium, show
no dramatic discrepancy below $2\,$Mpc~\cite{allen} if normal
gravity is assumed. On larger scales, there is no possible test on
the Poisson equation (although other aspects of GR can be tested,
see~\cite{dduality}) but by the mechanism of structure formation
through the development of gravitational instabilities. This is
the object of this letter.

Following \cite{UBbc1} we simply assume that the gravitational
force is changed at large-scale. Other possible modifications of
the large-scale gravity have been considered
in~\cite{othergravity}. We assume in particular that the
background metric is that of a Friedmann-Lema{\^\i}tre metric with an
expansion factor $a(z)$ having a redshift dependence assumed to be
that of a Univers with a cosmological constant. Note that it does
not mean the Friedmann equations is assumed to be correct, e.g.
that there is a dark energy component, but that we take the
evidences for a non-zero cosmological constant at face value. The
latter being mainly of cosmographical nature, e.g. they come from
the redshift dependence of the expansion parameter and quantities
that are directly related to it such as angular or luminosity
distances, they do not provide us with an actual weighing of the
matter/energy content of the Universe.

As a further consequence, in what follows, as long as we are
dealing with subhorizon scales, we can take the metric to be of
the form,
\begin{equation}
\dd s^2=-(1-2\Phi)\dd
t^2+a^2(1+2\Phi)\left(\dd\chi^2+q^2(\chi)\dd\Omega^2 \right),
\end{equation}
where $t$ is the cosmic time, $a(t)$ the scale factor, $\chi$ the
comoving radial coordinate, $\dd\Omega^2$ the unit solid angle and
$q(\chi)= (\sin\chi,\chi,{\rm sinh}\chi)$ according to the
curvature of the spatial sections.  In a Newtonian theory of
gravity, $\Phi$ is the Newtonian potential $\Phi_N$ determined by
the Poisson equation
\begin{equation}\label{poisson}
\Delta\Phi_N=4\pi G\rho a^2\delta
\end{equation}
where $G$ is the Newton constant and $\Delta$ the three
dimensional Laplacian in comoving coordinates, $\rho$ the
background energy density and $\delta\equiv\delta\rho/\rho$ is the
density contrast. If the Newton law is violated above a given
scale $r_s$ then we have to change Eq. (\ref{poisson}) and the
force between two masses distant of $r$ derives from $\Phi=\Phi_N
f(r/r_s)$  where $f(x)\rightarrow1$ when $x\ll1$.  This
encompasses for instance the potential considered
in~\cite{GRS,binetruy00} for which $f(x)=1/(1+x)$ (in that case
$f\propto1/x$ and 5D gravity is recovered at large distance).
Using (\ref{poisson}) it leads, with ${\bf r}=a{\bf x}$, to
\begin{equation}\label{ansatz}
\Phi({\bf x})=-G\rho a^2\int \dd^3{\bf x'}\frac{\delta({\bf
x'})}{|{\bf x}-{\bf x'}|}f\left(\frac{|{\bf x}-{\bf
x'}|}{x_s}\right),
\end{equation}
which, making use of $\Delta[f(x)/x]=-4\pi\delta^{(3)}({\bf
x})+f_s(x/x_s)$ with $f_s(x/x_s)\equiv(\partial_x^2f)/x$, gives
\begin{equation}\label{modNewt}
\Delta\Phi=\Delta\Phi_N-G\rho a^2\int\dd^3{\bf x'}\delta({\bf
x'}+{\bf x})f_s(x'/x_s).
\end{equation}

The motion equations of the field are better handled in Fourier
space. In Fourier space, Eq. (\ref{modNewt}) reads
\begin{equation}\label{loc}
-k^2\widehat\Phi(k)=4\pi G\rho a^2\widehat\delta(k)f_c(kr_s)
\end{equation}
where $f_c(k\,r_s)\equiv1-2\pi^2\,f_s(kr_s)$, $f_s(kr_s)$ being
the Fourier transform of $f_s(r/r_s)$. It is unity for large
values of $k\,r_s$, it behaves like $k\,r_s$ for small.

To explore the physics of large-scale structure growth the now
standard method~\cite{pert} is to expand the local density field
with respect to the initial density field. Such an expansion can
be made either in real space or in Fourier space,
\begin{equation}
\delta(\vk,t)=\delta^{(1)}(\vk,t)+\delta^{(2)}(\vk,t)+\dots
\end{equation}
where $\delta^{(1)}(\vk)$ is the linear density field,
$\delta^{(2)}(\vk)$ is quadratic in the linear density field, etc.
The motion equation for the density field can be solved order by
order and the resulting expansion can be used to infer the
evolution of the stochastic properties of the local scalar field,
whether it is the local density field or the gravitational
potential. More quantitatively for any stochastic field $X$
(statistically homogeneous and isotropic) we define its power
spectrum ${\cal P}_X$ by
\begin{equation}\label{defP}
\langle \widehat X(\vk)\widehat X(\vk') \rangle\equiv {\cal
P}_X(k)\delta^{(3)}(\vk+\vk')
\end{equation}
where $\delta^{(3)}$ is the Dirac distribution, $\widehat X$ the
Fourier transform of $X$ and the brackets refer to ensemble
averages~\cite{pert} over the stochastic process that gave birth
to $X$. Note that the poisson equation then implies,
\begin{equation}\label{poisson_fourier}
{\cal P}_{\Delta\Phi_N}(k)=\left(4\pi G\rho a^2\right)^2 {\cal
P}_\delta(k).
\end{equation}
It is usually assumed that the initial metric fluctuation are
Gaussian distributed. In the early stage of the dynamics it is
therefore enough to consider the power spectrum for the scalar
fields to fully characterize them. The subsequent evolution of the
field however is bound to induce non-Gaussian effects. The second
order $\delta^{(2)}(\vk,t)$ in particular contains mode coupling
terms that induce non-zero third moments (whether it is in real or
in Fourier space). One way to quantify those effects is to
introduce the field bispectrum, $\mB_X(\vk_1,\vk_2,\vk_3)$ defined
as,
\begin{equation}
\langle \widehat X(\vk_1)\widehat X(\vk_2) \widehat
X(\vk_3)\rangle\equiv {\cal
B}_X(\vk_1,\vk_2,\vk_3)\delta^{(3)}(\vk_1+\vk_2+\vk_3).
\end{equation}
In cosmology it is common to define the reduced bispectrum as,
\begin{equation}\label{Qdef}
Q_X(\vk_1,\vk_2,\vk_3)=\frac{\mB_X(\vk_1,\vk_2,\vk_3)}{\mP_X(k_1)\mP_X(k_2)+sym.}
\end{equation}
the reason being that $Q_{\delta}$ turns out to be roughly scale
independent.

With modified gravity laws we can already note that,
\begin{equation}
{\cal P}_{\Delta\Phi}(k)=\left(4\pi G\rho a^2\right)^2{\cal
P}_{\delta}(k)f_c(k\,r_s)^2
\end{equation}
and
\begin{equation}\label{Qscaling}
Q_{\Delta\Phi}(k)=\frac{Q_{\delta}(k)}{f_c(k\,r_s)}
\end{equation}
for equilateral wave vector configurations.

As a consequence, a simple way to test the validity of the Newton
law is to compare power spectra of the density field and that of
$\Delta\Phi$ to see if they follow the same scale dependence. That
was the point made in \cite{UBbc1}. This is possible however only
if one can measure $\delta$ and $\Phi$ independently. That
supposes in particular that galaxy counts can give a reliable
account of the mass distribution at large-scale, which as some
level is always questionable due to the unavoidable presence of
biasing effects\footnote{Although this limitation might not be
that critical at large enough scale~\cite{bias}.}  Note that
constraints on modified gravity law can also be obtained from the
shape of the power spectrum alone~\cite{sealfon} but such
constraints rely on assumptions on the shape of the spectrum of
the primordial metric fluctuations.

The aim of this letter is to show that it is possible to test
departure from standard gravity from cosmic shear survey alone
taking advantage of the behavior of the reduced bispectrum.

As long as we have not entered too much into the nonlinear regime,
the amplitude and shape of the power spectrum is determined by the
growth rate of the linear term, $\delta^{(1)}(\vk,t)$. In such
model, and contrary to cosmological models with standard gravity,
the growth rate is scale dependence. More precisely we have,
\begin{equation}
\delta^{(1)}(\vk,t)=D_k(t)\deltal(\vk)
\end{equation}
where $D_k(t)$ is the linear growing rate of mode $k$ that
satisfies the equation~\cite{pert},
\begin{equation}
\ddot D_k(t)+2H\dot D_k(t)= \frac{3}{2}H^2\,\Omega(t)\,f_c\left(k
r_s/a(t)\right)D_k(t).
\end{equation}
The solution of this equation have already been explored
in~\cite{UBbc1}. It leads to a slowing of the growth rate for the
scales beyond $r_c$.

For the 2nd order terms the structure of the conservation and
motion equations leads to the following functional form (see also
the approach adopted in~\cite{DPGgrav}),
\begin{eqnarray}
\delta^{(2)}({\vk})&=&\int\d^3\vk_1\int\d^3\vk_2\,{\delta_{\rm
Dirac}(\vk_1+\vk_2)\over (2\pi)^{3/2}}\left[
F_0(\vk_1,\vk_2)\deltal(\vk_1)\deltal(\vk_2)+\right.\nonumber\\
&&\left.\hspace{1cm}
F_1(\vk_1,\vk_2)\deltal(\vk_1)\deltal(\vk_2)
\frac{\vk_1.\vk_2}{k_1 k_2 }+
F_2(\vk_1,\vk_2)\deltal(\vk_1)\deltal(\vk_2)
\left(\frac{(\vk_1.\vk_2)^2}{k_1^2
k_2^2}-\frac{1}{3}\right)\right]
\end{eqnarray}
where the coefficients $F_0$, $F_1$ and $F_2$ represent the
amplitude of respectively the monopole, dipole and quadrupole
terms that appear of the expression of the 2nd order local density
field. For cosmological models with a standard gravity, $F_0$,
$F_1$ and $F_2$ are scale independent. Moreover they appear to
behave like $D^2(t)$ with a fixed coefficient that depends
extremely weakly on the cosmological parameters.

Lest us now explore the case of a modified gravity dynamics. The
continuity equation imposes that $\mg\delta^{(2)}\md=0$ which
implies that $F_0-F_1+2/3 F_2\equiv 0$. The behavior of the 2nd
order local density field will then be obtained from the evolution
equation for these 2 functions,

\begin{eqnarray}\label{evodelta}
\ddot F_0-2H\dot F_0- \frac{3}{2}H^2\,\Omega(t)\,f_c\left(
\vert\vk_1+\vk_2\vert\frac{r_s}{a(t)}\right)F_0&=& \frac{4}{3}\dot
D_{k_1}\dot D_{k_2}+\frac{3}{4}H^2 \left[
f_c\left(k_1\frac{r_s}{a(t)}\right)+f_c\left(
k_2\frac{r_s}{a(t)}\right)\right]D_{k_1}D_{k_2}\\
%\ddot F_1-2H\dot F_1- \frac{3}{2}H^2\,\Omega(t)\,f_c\left(
%\vert\vk_1+\vk_2\vert\frac{r_s}{a(t)}\right)F_1&=& 2\dot D
%_{k_1}\dot D_{k_2}+\frac{3}{4}H^2 \left[
%f_c\left(k_1\frac{r_s}{a(t)}\right)+f_c\left(
%k_2\frac{r_s}{a(t)}\right)\right]D_{k_1}D_{k_2}
\end{eqnarray}
and ${4/3}\dot D_{k_1}\dot D_{k_2}$ in the r.h.s. of the previous
equation is changed into $2\dot D_{k_1}\dot D_{k_2}$ for the
evolution equation of $F_1$.

These two equations have a very similar structure and can be
solved numerically for whichever background we consider. To get
the behavior of $F_0$ and $F_1$ in the following we assume we live
in a universe in which,
\begin{equation}\label{adez}
  a(t)=a_0\,\sinh\left[\frac{3H_0\,t}{2}
  \left(1-\Omega_0\right)^{1/2}\right]
\end{equation}
which corresponds to the expansion factor behavior for a universe
for a given value of $\Omega_0$ and no curvature (in the following
we do the calculations with $\Omega_0=0.3$). Note that we found
that the results described in the following are very weakly
dependent on the background behavior.

\begin{figure}
  % Requires \usepackage{psfig}
  \centerline{ \psfig{figure=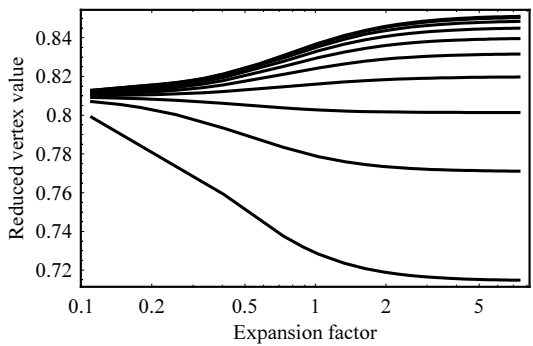,width=7cm}}
  \caption{Time dependence of $F_0(\vk_1,\vk_2)/D_{k_1}/D_{k_2}$
  for $k_1=k_2$ and for different value of the relative angle between $\vk_1$
  and $\vk_2$. This angle varies from $0$ to $\pi$ from top to bottom lines. }
  \label{nu2}
\end{figure}

Compared to the standard gravity case, $F_0$, $F_1$ and $F_2$ now
depend on the wave vectors through both their length and their
relative angle. It introduces further dependence of the bispectrum
on the triangle configuration. As an illustration we present on
Fig. \ref{nu2} the behavior of $F_0(\vk_1,\vk_2)/D_{k_1}/D_{k_2}$
as a function of time and for different angle between the two
wavevectors (of the same length). The angular modulation is
exhibited at late time (at early time this ratio is $17/21$
irrespectively of the angle). It is found to be actually modest
given the precision with which such a quantity can be measured.
Similar results can be found for $F_1$.

\begin{figure}
  % Requires \usepackage{psfig}
  \centerline{ \psfig{figure=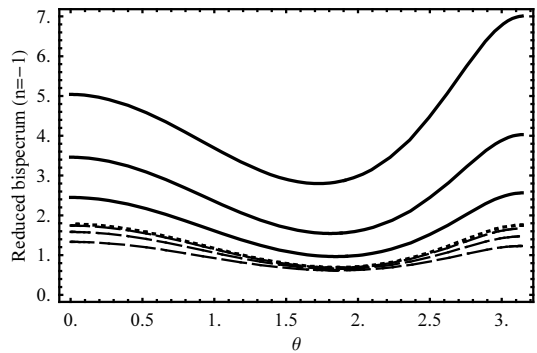,width=7cm}}
  \caption{Shape of the bispectrum for the density field (long dashed lines) and the
  potential Lagrangian (solid lines). The computation has
  been made for $k_1=r_s/10$ and $k_2=2 r_s/10$
  and assuming $P(k)\sim k^{-1}$ for the density field.}
  \label{bispectrumCM}
\end{figure}

The resulting behavior of the bispectrum are show on Fig.
\ref{bispectrumCM}. It shows the time evolution of the reduced
bispectrum $Q$ as defined in Eq. (\ref{Qdef}) for either the
density field or the potential Laplacian compared to the standard
gravity case. The density bispectrum can be measured for instance
in galaxy catalogues. It has been done in particular in the PSCz
galaxy survey~\cite{PSCz}. It does not show however strong
dependence on the modified gravity effects. This is not the case
for the Laplacian field whose reduced skewness is amplified by
modified gravity effects. That was indeed to be expected from the
scaling relation shown in Eq. (\ref{Qscaling}) which suggests that
such a quantity is expected to grow at large scale.

\begin{figure}
  % Requires \usepackage{psfig}
  \centerline{ \psfig{figure=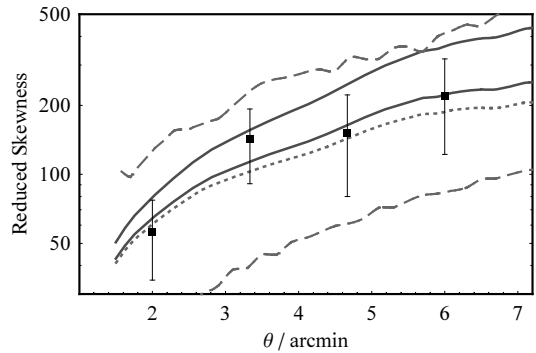,width=7cm}}
  \caption{The measured reduced shear skewness (square with error bars)
  compared to cosmological models with standard gravity (dotted line for
  $\Lambda$-CDM, upper dashed line for open CDM and lower dashed line
  for standard CDM) or with modified gravity for $\Lambda$ CDM,
  $r_s=10h^{-1}$ Mpc (lower solid line) and $r_s=2h^{-1}$ Mpc (upper solid line).}
  \label{xi3}
\end{figure}

It happens that quantities directly related to the reduced third
moment of the potential Laplacian  have been measured in cosmic
shear catalogues~\cite{xi3detec}. The three-point correlation of
the shear field (or rather some geometrical average of it) has
been detected with a significant confidence level (through a
rather elaborate procedure) at different angular
scales~\cite{xi3method}. If one applies the scaling law one
expects for modified gravity to the predictions of standard
gravity (in a $\Lambda$ dominated universe) it is possible to put
constraints on $r_s$. From Fig. \ref{xi3} it appears that current
data already disfavor values of $r_s$ below $2 h^{-1}$ Mpc
although it is probably impossible to put a definitive statement
from this data set. The constraints on the scale at which modified
gravity intervenes are however strongly dependent on the modelling
of the observations, and on the detailed shape of the 4D to 5D
transition.

Cosmic shear surveys appear nonetheless very attractive to test
the validity of the Einstein gravity law at cosmological scales.
The fact such surveys provide us with genuine potential
distribution at large scale makes them precious probes of the
gravitational instability mechanisms in general. Undoubtedly, much
stringent constraints should be obtainable from the coming cosmic
shear surveys such as the CFHTLS or on the long term satellite
missions like JDEM/SNAP.

Note that should such a modified gravity be the reason of the
observed evidences of an accelerated universe $r_s$ would be
expected to be the order of the Hubble size. Obtaining constraints
on $r_s$ of this order form large-scale surveys may not be as
hopeless at it seems. For the potential shape derived
in~\cite{Deffayet1} parameterized by a scale $r_c$ whose estimated
value is  $r_c=1.2\,c/H_0$~\cite{Deffayet2}, the amplification of
the three-point correlation is not as small for the available
scales as the simplified model used here suggests. At 10
$h^{-1}$Mpc scale the effect would be at percent level; at 100
$h^{-1}$Mpc scale, it would reach 7\% (a valid theory of gravity
has however to be done in those models). If this approach reveals
the only way to distinguish between quintessence type and modified
gravity models to account for the dark energy evidences, it calls
for the realization of very large - possibly whole sky - cosmic
shear surveys.

%%%%%%%%%%%%%%%%%%%%%%%%%%%%%%%%%%%%%%%%%%%%%%%%%%%%%%%%%%%%%%%

%%%%%%%%%%%%%%%%%%%%%%%%%%%%%%%%%%%%%%%%%%%%%%%%%%%%%%%%%%%%%%%
\end{document}